# Constructing Hyperbolic Metamaterials with Arbitrary Medium

Li-Zheng Yin, Feng-Yuan Han, Jin Zhao, Di Wang, Tie-Jun Huang and Pu-Kun Liu*

State Key Laboratory of Advanced Optical Communication Systems and Networks,

Department of Electronics, Peking University, Beijing, 100871, China

*Corresponding author: pkliu@pku.edu.cn

## Abstract

Recent advances in hyperbolic metamaterials have spurred many breakthroughs in the field of manipulating light propagation. However, the unusual electromagnetic properties also put extremely high demands on its compositional materials. Limited by the finite relative permittivity of the natural materials, the effective permittivity of the constructed hyperbolic metamaterials is also confined to a narrow range. Here, based on the proposed concept of structure-induced spoof surface plasmon, we prove that arbitrary materials can be selected to construct the hyperbolic metamaterials with independent relative effective permittivity components. Besides, the theoretical achievable ranges of the relative effective permittivity components are unlimited. As proofs of the method, three novel hyperbolic metamaterials are designed with their functionalities validated numerically and experimentally by specified directional propagation. To further illustrate the superiority of the method, an all-metal low-loss hyperbolic metamaterial filled with air is proposed and demonstrated. The proposed methodology effectively reduces the design requirement for hyperbolic metamaterials and provides new ideas for the scenarios where large permittivity coverage is needed such as microwave and terahertz focus, super-resolution imaging, electromagnetic cloaking, and so on.

Hyperbolic metamaterials(HMMs) are a class of anisotropic artificial media whose equal dispersion contours(EFCs) are hyperbola. This unusual property is determined by their effective electric or magnetic tensors where one of the principal components of either their permittivity or permeability tensors is opposite in sign to the other two principal components. In this work, we only focus on the electric HMMs whose relative permeability tensors degenerate to a constant. Different from the natural anisotropic materials, due to the openness of the hyperbola, HMMs can support the electromagnetic modes with arbitrarily large wave vector that are otherwise evanescent waves in free space. These unique properties make the HMMs ubiquitous in different areas ranging from subwavelength resolution imaging, [1-5] negative refraction, [6-9] focusing, [10-13] thermal emission engineering, [14-15] spontaneous emission enhancement, [16-19] and so on. In the optical and infrared regimes, the periodic stacking of subwavelength metal-dielectric structures is the most common way to obtain hyperbolic dispersion. The diagonal components of the effective permittivity tensors with opposite signs can be independently tuned by adjusting the permittivity of the dielectric and the metal, and their thickness ratio. [20-22] This effect is attributed to the continuous excitation of surface plasmon polaritons (SPPs) at each metal-dielectric interface. Besides, substituting the metal with the plasmonic graphene, tunable HMMs can be obtained by adding external voltage to the graphene layers. [23-26] It should be noted that this tunability is only limited to the negative permittivity components, and the positive effective permittivity components are equal to that of the dielectric.

However, in the microwave and terahertz regimes, the methods mentioned above cannot function because almost all of the natural plasmonic materials, within their limited choices, are only available in the infrared and visible wavelengths. [27-28] In 2016, Nader Engheta *et.al.* propose a concept of "effective medium" by filling the waveguides with specific materials. In this way, surface plasmon can be induced by pairing the effective positive and negative materials. [29-30] Based on this concept, hyperbolic dispersion can be obtained by stacking the effective positive and negative medium periodically. This method successfully gets rid of the dependence on the conventional plasmonic materials in designing HMMs. However, to construct a HMM with a specific effective permittivity tensor, one should also search for the practical materials with the specified permittivity laboriously. [31-33] Recently, our group proposes the concept of equivalent graphene metasurfaces (EGMs), a two-dimension (2D) structure that can mimic the property of plasmonic graphene in the microwave and terahertz regimes. [34] HMMs can be constructed by stacking the EGMs and dielectric periodically. This method effectively breaks the restrictions on filling materials. However, it faces the same problem as the graphene HMMs that the positive relative effective permittivity components cannot be adjusted.

In this work, we propose the concept of structure-induced spoof surface plasmon(SISSP) supported by waveguide plasmonic metasurfaces(WPMs) and prove that its propagation and decay constants can be independently adjusted merely by the structure parameters. Based on this concept, we demonstrate that hyperbolic dispersion can be obtained by stacking the WPMs periodically. Compared with the existing works, the proposed methodology completely breaks the barriers of materials, and arbitrary materials can be selected to construct the HMMs with independent effective permittivity components. Besides, the theoretical achievable range of the relative effective permittivity is unlimited. To demonstrate the advantages of large permittivity coverage, we design three HMMs whose relative effective permittivity components meet the conditions $\varepsilon_\perp \ll |\varepsilon_{//}|$, $\varepsilon_\perp = |\varepsilon_{//}|$, and $\varepsilon_\perp \gg |\varepsilon_{//}|$, respectively. For sake of verifying the material-independent property, here, the relative permittivity of the filling materials of all the HMMs is $\varepsilon_r = 4.4 - 0.01i$. The experiments performed in the microwave regime matched well with the theoretical and simulation results, which demonstrates the validity of the proposed method. To further explore its potential in low-loss scenarios, we design an air-medium HMM whose damping in time has an obvious reduction compared with the normal ones. This superiority is more evident and valuable in the terahertz range where the loss of dielectric is quite higher. Finally, in the Discussion and Conclusion, we discuss the practical applications and advantages of the proposed method in designing epsilon-near-zero (ENZ) materials and super-resolution lenses.

## Results and Discussion

### Concept of Structure-Induced Spoof Surface Plasmon

The structure-induced or the geometry-induced dispersion whose characteristic is different from the filling materials is due to the role of the geometry of the structures on wave propagation. The parallel-plate waveguides(PPWs) and the rectangular waveguides(RWs) filled with the specific materials, which are also called waveguide metamaterials, are the most common structures to realize the structure-induced dispersion. For the convenience of analysis, the influence of the PPWs or the RWs can be reflected on the filling materials by expressing its relative effective permittivity with $\varepsilon_e = \varepsilon_r - \lambda_0^2/4a^2$, where $\varepsilon_r$ is the real relative permittivity of the materials, $\lambda_0$ is the wavelength in free space, and $a$ is the interval between the waveguide walls. This equation tells us arbitrary relative effective permittivity smaller than $\varepsilon_r$ can be obtained by adjusting $a$. In this way, the effective surface plasmon polariton(ESPPs) can be formed by pairing the waveguide metamaterials with positive and negative relative effective permittivity. [29-30, 35-37] The corresponding propagation constant and decay constants can also be obtained with

$$\beta = k_0 \sqrt{\frac{(\varepsilon_{r1} - \lambda_0^2/4a^2)(\varepsilon_{r2} - \lambda_0^2/4a^2)}{\varepsilon_{r1} + \varepsilon_{r2} - \lambda_0^2/2a^2}}, \qquad (1)$$

$$k_i = k_0(\varepsilon_{ri} - \lambda_0^2/4a^2)\sqrt{\frac{-1}{\varepsilon_{r1} + \varepsilon_{r2} - \lambda_0^2/2a^2}}, \qquad (2)$$

respectively, where $i = 1, 2$ represent the different materials, and $k_0$ is the wavevector in free space. Although defying the limitation of the plasmonic materials whose choices are quite limited, according to eqs 1 and 2, to realize the ESPPs with the specific $\beta$ and $k_i$, one should also search for the practical materials with the right permittivity laboriously. It can also be interpreted that, for the specified $\varepsilon_{r1}$ and $\varepsilon_{r2}$, the only one freedom $a$ is not adequate to tune the $\beta$ and $k_i$ independently. This property puts a very high demand on the filling materials in supporting the specified ESPPs. The situations are the same when we design the HMMs with layered effective positive-negative

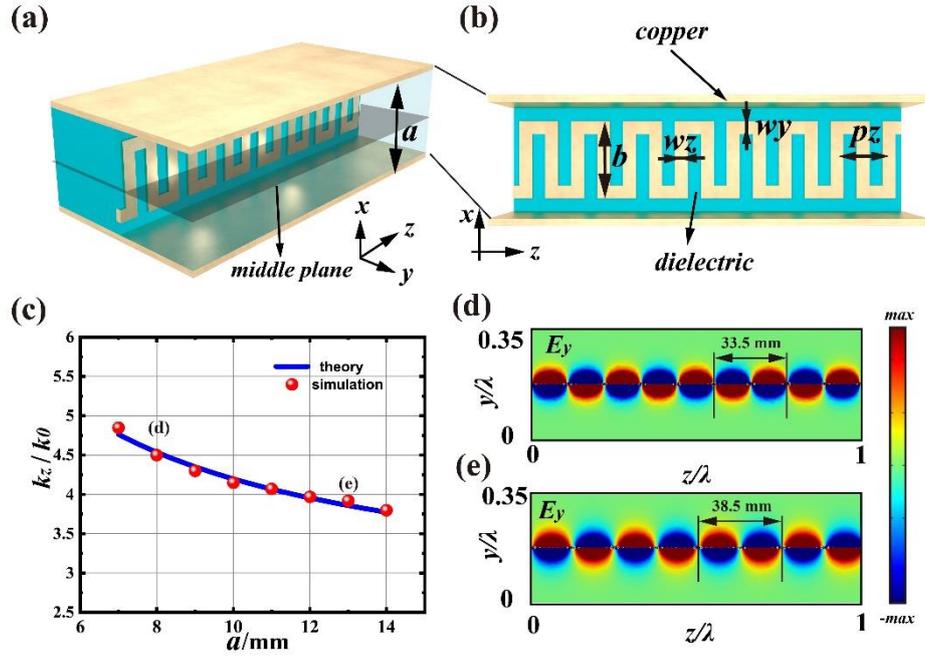

Figure 1. (a) The schematic and (b) the detailed parameters of the proposed WPM that supports the SISSP. (c) The theoretical and simulated propagation constants of the SISSP versus different interval $a$ under the effective conductance of the PM $\sigma_m = -14.7i$ mS. (d) and (e) are $E_y$ distributions on the middle plane in (a) with $a$ = 8 mm and 13 mm, respectively.

materials. Limited by the finite permittivity of the natural materials, the effective permittivity of the constructed HMMs is also confined to a narrow range. [31-33] To reduce the dependence on the relative permittivity of the dielectric, in this work, we propose a novel waveguide metamaterial with the plasmonic metasurfaces(PMs) embedded in the PPWs vertically. As illustrated by Figure 1a, In this work, we introduce the meander lines as a specified realization of PMs whose effective conductance and dispersion characteristics are analyzed in detail in Supporting Information. Figure 1b illustrates its shape and structure parameters. Compared with single PMs, the proposed WPMs can effectively expand the tuning range of the dispersion characteristic and even realize some extreme dispersion relation that natural materials cannot reach. To demonstrate these merits, we deduce its Eigen field distribution with the method of Borgnis potential function. Here, we only consider the TM mode (the magnetic field components lay in the $xy$ plane) so the Borgnis potential functions $V = 0$ and $U$ of the structure has the

form

$$U = \begin{cases} Ae^{-jk_z z}e^{jk_y y}\cos(k_{mx}x), & y > 0 \\ Ae^{-jk_z z}e^{-jk_y y}\cos(k_{mx}x), & y < 0 \end{cases} \quad (3)$$

where $k_z$ and $k_y$ are the propagation constant and wave vector components in $y$ directions, respectively. $k_{mx} = m\pi / a$ represents the lateral metal-boundary-induced wave vector, where $m$ is an arbitrary integer and $a$ is the interval of the PPW. The electric and magnetic field components above the plasmonic metasurface can be calculated according to Borgnis potential functions U

$$E_{1x} = \frac{\partial^2 U}{\partial x \partial y} = -Ajk_{mx}k_y e^{-jk_z z}e^{jk_y y}\sin(k_{mx}x), \quad (4a)$$

$$E_{1y} = \frac{\partial^2 U}{\partial y \partial z} = Ak_y k_z e^{-jk_z z}e^{jk_y y}\cos(k_{mx}x), \quad (4b)$$

$$E_{1z} = \frac{\partial^2 U}{\partial z^2} + \varepsilon_r k_0^2 U = A(k_{mx}^2 + k_y^2)e^{-jk_z z}e^{jk_y y}\cos(k_{mx}x), \quad (4c)$$

$$H_{1x} = j\omega\varepsilon_0\varepsilon_r \frac{\partial U}{\partial y} = -A\omega\varepsilon_0\varepsilon_r k_y e^{-jk_z z}e^{jk_y y}\cos(k_{mx}x), \quad (4d)$$

$$H_{1y} = -j\omega\varepsilon_0\varepsilon_r \frac{\partial U}{\partial x} = Aj\omega\varepsilon_0\varepsilon_r k_{mx} e^{-jk_z z}e^{jk_y y}\sin(k_{mx}x). \quad \text{for } y > 0 \quad (4e)$$

Similarly, the corresponding field distribution below the plasmonic metasurface can also be obtained

$$E_{2x} = \frac{\partial^2 U}{\partial x \partial y} = Ajk_{mx}k_y e^{-jk_z z}e^{-jk_y y}\sin(k_{mx}x), \quad (5a)$$

$$E_{2y} = \frac{\partial^2 U}{\partial y \partial z} = -Ak_y k_z e^{-jk_z z}e^{-jk_y y}\cos(k_{mx}x), \quad (5b)$$

$$E_{2z} = \frac{\partial^2 U}{\partial z^2} + \varepsilon_r k_0^2 U = A(k_{mx}^2 + k_y^2)e^{-jk_z z}e^{-jk_y y}\cos(k_{mx}x), \quad (5c)$$

$$H_{2x} = j\omega\varepsilon_0\varepsilon_r \frac{\partial U}{\partial y} = A\omega\varepsilon_0\varepsilon_r k_y e^{-jk_z z}e^{-jk_y y}\cos(k_{mx}x), \quad (5d)$$

$$H_{2y} = -j\omega\varepsilon_0\varepsilon_r \frac{\partial U}{\partial x} = Aj\omega\varepsilon_0\varepsilon_r k_{mx} e^{-jk_z z} e^{-jk_y y} \sin(k_{mx}x), \quad \text{for } y < 0 \tag{5e}$$

where $k_y = \sqrt{\varepsilon_r k_0^2 - k_x^2 - k_z^2}$. When we model the PMs as an ultrathin sheet with effective conductance $\sigma_m$ ($\omega$, $\varepsilon_r$),[34, 38] the boundary conditions between the two regions can be expressed as

$$E_{1x} = E_{2x}|_{y=0}, \tag{6a}$$

$$E_{1z} = E_{2z}|_{y=0}, \tag{6b}$$

$$H_{2x} - H_{1x} = \sigma_m(\omega, \varepsilon_r) E_z|_{y=0}. \tag{6c}$$

The practical effective conductance $\sigma_m$ ($\omega$, $\varepsilon_r$) can be numerically calculated through finite element method(FEM) simulation (See Supporting Information for the detailed analysis). By substituting eqs 2 and 3 into eq 4, the dispersion relation of the proposed WPM is obtained

$$2k_y = \frac{\sigma(\omega,\varepsilon_r)}{\omega\varepsilon_0\varepsilon_r}(k_{mx}^2 + k_y^2) \tag{7}$$

Assuming $k_{y0} = 2\omega\varepsilon_0\varepsilon_r / \sigma_m(\omega, \varepsilon_r)$, the wave vector component $k_y$ whose real part represents the decay constant can be solved from eq 5

$$k_y = \frac{k_{y0} + \sqrt{k_{y0}^2 + 4k_{mx}^2}}{2} \tag{8}$$

The defined constant $k_{y0}$ represents the Eigen lateral wave vector component of the PMs without the PPWs. It should be noted that the ideal effective conductance of the PMs is approximate an imaginary number because the real part is small enough to omit. Therefore, the proposed WPMs only support evanescent waves. From eq 8 we can find that the existence of the metal boundary effectively increases the lateral wave vector component, that is, increases the electric field confinement. Besides, the propagation constant $k_z$ can also be obtained

$$k_z = \sqrt{\varepsilon_r k_0^2 - k_y^2 - k_{mx}^2} = \sqrt{\varepsilon_r k_0^2 - 2k_{mx}^2 - \frac{k_{y0}^2 + k_{y0}\sqrt{k_{y0}^2 + 4k_{mx}^2}}{2}}. \tag{9}$$

Equation 9 tells us the propagation constant of the SSPs can be adjusted not only by the effective conductance of the PMs but also by the interval of the PPWs. Compared with the ESPPs supported by the conventional waveguide metamaterials, the SISSP in the WPMs provides us one more freedom, i.e. effective conductance $\sigma_m$ ($\omega$, $\varepsilon_r$), to tune its dispersion characteristic. Although $\sigma_m$ is the functions of $\varepsilon_r$, the practical value of the effective conductance is dominantly controlled by the shape and structure parameters of the PMs (see Supporting Information for the detailed analysis). In this case, the propagation constant $k_z$ and the lateral wave vector component $k_y$ can be independently controlled through adjusting the interval $a$ and the effective conductance $\sigma_m$ ($\omega$, $\varepsilon_r$), which effectively reduces the dependence on the relative permittivity of the dielectric. To demonstrate the properties of the SISSP, we calculate its theoretical and numerical propagation constants with $a$ ranging from 7 mm to 14 mm at 2 GHz, and plot the results in Figure 1c. Here, the structure parameters of the embedded PM with the effective conductance $\sigma = -14.7i$ mS are $w_y = w_z = 1$ mm, $b = 6$ mm, $\varepsilon_r = 4.4 - 0.01i$, and $p_z = 4$ mm. As illustrated by Figure 1c, the consistency between the theoretical and numerical propagation constants effectively validates the effectiveness of the proposed method. To visualize the practical effect, Figures 1d and 1e show the electric field $E_y$ distribution on the middle plane of Figure 1a with $a = 8$ mm and 13 mm, respectively. The superior characteristics of the SISSP make it hold great promise in designing microwave and terahertz circuits, metamaterials, and so on.

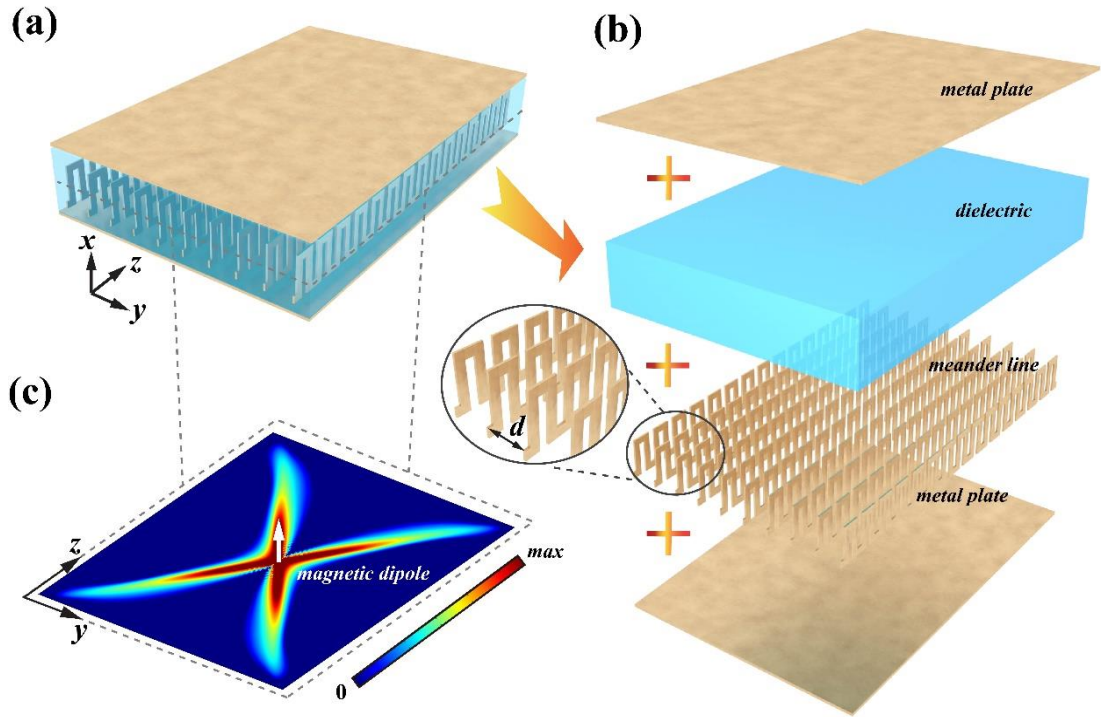

Figure 2. (a) Schematic of the proposed anisotropic metamaterial constructed by multilayer WPMs. (b) The complete construction of the structure, where the inset displays the detail of the meander lines. (c) The typical hyperbolic field distribution excited by a magnetic dipole on the middle plane of the metamaterials.

## Construct HMMs with SISSP

To further explore the potential of the SISSP, we construct the metamaterials by stacking the WPMs periodically, as illustrated by Figure 2a. Figure 2b illustrates the complete construction of the structure, where the inset displays the details of the meander-line WPMs. The distance between the adjacent WPMs is $d$. To obtain the dispersion relation of the constructed metamaterials, we write the transfer matrix of the WPM

$$T_1 = \begin{bmatrix} 1 & 0 \\ \sigma_m & 1 \end{bmatrix}. \tag{10}$$

Similarly, the transfer matrix of the dielectric with thickness $d$ has the form

$$T_d = \begin{bmatrix} \cos(k_y d) & jZ_d \sin(k_y d) \\ \dfrac{j}{Z_d} \sin(k_y d) & \cos(k_y d) \end{bmatrix}, \tag{11}$$

where $Z_d$ represents the wave impedance and is relevant to the modes supported by the WPMs. $k_y = \sqrt{\varepsilon_r k_0^2 - k_x^2 - k_z^2}$, where $\varepsilon_r$ is the real relative permittivity of the filling materials. In this case, the transfer matrix of a complete WPM unit can be determined by multiplying eqs 10 and 11

$$T_{unit} = T_\sigma T_d = \begin{bmatrix} \cos(k_y d) & jZ_d \sin(k_y d) \\ \dfrac{j}{Z_d} \sin(k_y d) + \sigma_m \cos(k_y d) & j\sigma_m Z_d \sin(k_y d) + \cos(k_y d) \end{bmatrix}. \tag{12}$$

Assuming the tangential electromagnetic wave(EMW) at $y$ has the form $\begin{bmatrix} E_t(y) \\ H_t(y) \end{bmatrix}$, according to the Bloch theory, the tangential components at $y + d$ which have travelled over one space period, can be written as

$$\begin{bmatrix} E_t(y+d) \\ H_t(y+d) \end{bmatrix} = e^{-jk_\perp d} \begin{bmatrix} E_t(y) \\ H_t(y) \end{bmatrix}, \quad (13)$$

where $k_\perp$ represents the Bloch wave vector. Meanwhile, the tangential components at $y + d$ can also be calculated by Transfer Matrix Method

$$\begin{bmatrix} E_t(y+d) \\ H_t(y+d) \end{bmatrix} = T_{unit} \begin{bmatrix} E_t(y) \\ H_t(y) \end{bmatrix}. \quad (14)$$

Combining eqs 11 and 12, the dispersion relation of metamaterial constructed by multilayer WPMs can be thus determined from the solution of the equation

$$\det \left| T_{unit} - e^{-jk_\perp d} I \right| = 0, \quad (15)$$

where **I** represents the identity matrix. Substituting eq 12 into eq 15, we can obtain

$$\cos(k_\perp d) = \cos(k_y d) + j\frac{1}{2}\sigma_m Z_d \sin(k_y d). \quad (16)$$

Equation 16 represents the theoretical dispersion relation of the constructed metamaterials. For simplicity, in this work, we only consider the condition in which the distance $d$ between the adjacent WPMs is deep subwavelength so the inequation $k_z d \ll 1$ and $k_\perp d \ll 1$ always hold. Under these preconditions, we substitute the equivalent infinitesimal $\cos(k_y d) = 1 - \frac{1}{2}(k_y d)^2$, $\cos(k_\perp d) = 1 - \frac{1}{2}(k_\perp d)^2$, and $\sin(k_y d) = k_y d$ into eq 16 and the simplified dispersion relation arrives at

$$k_\perp^2 = k_y^2 - j\frac{\sigma_m k_y Z_d}{d}. \quad (17)$$

Different from the conventional situation. Here, the wave impedance determined by the tangential components of the Eigen EMWs is

$$Z_d = \frac{E_z}{H_x} = \frac{k_{mx}^2 + k_y^2}{\omega \varepsilon_0 \varepsilon_r k_y}, \quad (18)$$

where $E_z$ and $H_x$ can be obtained from the Eigen field distribution eqs 4 and 5 of the WPMs. Defining $\varepsilon_e = \varepsilon_r - j\dfrac{\sigma_m(\omega,\varepsilon_r)}{\omega\varepsilon_0 d}$, the dispersion relation is finally simplified as

$$\frac{k_\perp^2}{\varepsilon_e - (\frac{\lambda}{2a})^2} + \frac{k_z^2}{\varepsilon_r - \frac{\varepsilon_r}{\varepsilon_e}(\frac{\lambda}{2a})^2} = k_0^2. \tag{19}$$

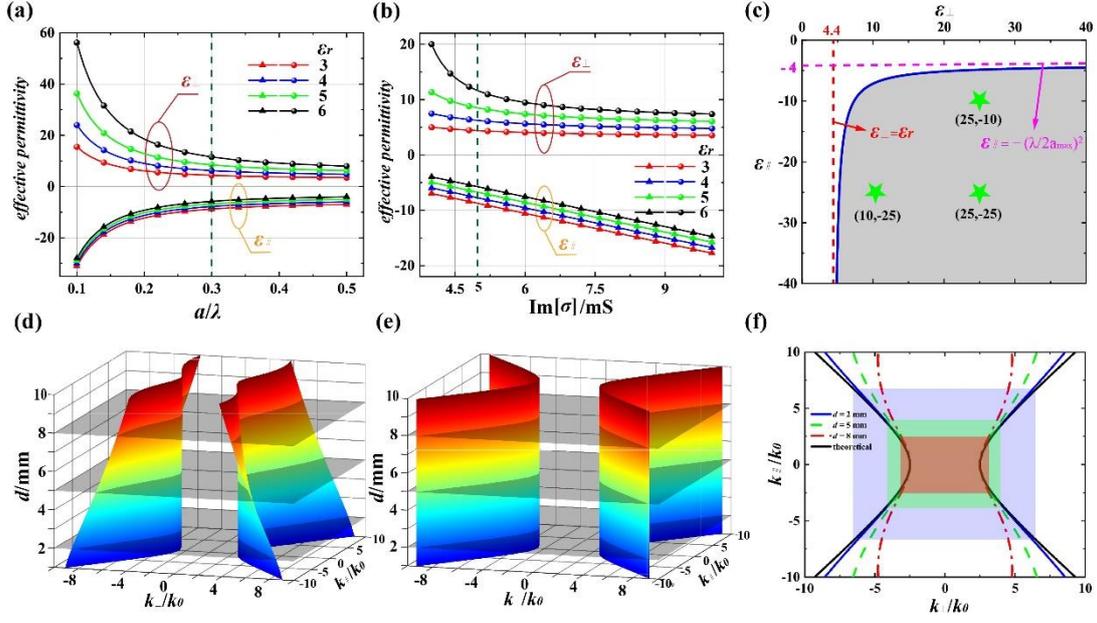

Figure 3. (a) and (b) show the effective permittivity of the HMMs with different $\sigma_m$, $\varepsilon_r$, and $a$. (a) is calculated under $\sigma_m = 5i$ mS, and (b) is calculated under $a = 0.3\ \lambda$. The dotted green lines in (a) and (b) represent the intersection position of the two figures. (c) The realizable area of the effective permittivity components of the proposed HMMs. (d) and (e) represent the practical and theoretical EFCs of the proposed HMM versus different $d$ with $\varepsilon_\perp = 6.2$, $\varepsilon_{//} = -7.8$. (f) The 2D EFCs corresponding to the cross sections in (d) and (e).

Therefore, the whole structure can be treated as a uniform anisotropic material with effective permittivity tensor

$$\hat{\varepsilon} = \begin{bmatrix} \varepsilon_\perp & 0 \\ 0 & \varepsilon_{//} \end{bmatrix}, \tag{20}$$

where $\varepsilon_{//} = \varepsilon_e - (\lambda/2a)^2$, and $\varepsilon_\perp = \varepsilon_r - (\lambda/2a)^2\ \varepsilon_r/\varepsilon_e$. This method opens up new routes in designing anisotropic metamaterials. In this work, we only focus on the HMMs which have broad practical applications in many areas. Figure 2c illustrates the typical magnetic field distribution of the HMMs excited by a magnetic dipole on the middle plane of the HMM in Figure 2a. As can be seen from eq 19, there are three key parameters that can determine the two effective permittivity components of the anisotropic metamaterials, i.e. the effective conductivity which is the ratio of the effective conductance to the distance of the adjacent WPMs $\sigma_m / d$, the real relative permittivity of the filling material $\varepsilon_r$, and the interval between the metal plate $a$. Therefore, for any specified $\varepsilon_r$, we can still design the HMMs with arbitrary and independent effective permittivity by only adjusting $\sigma_m / d$ and $a$ which can be determined by

$$a = \frac{\lambda}{2\sqrt{\varepsilon_{//}(\varepsilon_r - \varepsilon_\perp)/\varepsilon_\perp}}, \tag{21a}$$

$$\sigma_m / d = -j\omega\varepsilon_0\varepsilon_r(1 - \varepsilon_{//}/\varepsilon_\perp). \tag{21b}$$

This method effectively reduces the dependence on the materials in designing HMMs. In this way, arbitrary materials can be selected to construct HMMs. As a demonstration, we calculate the effective permittivity of the HMMs versus different $a$, $\sigma_m$, and $\varepsilon_r$. Figures 3a is calculated under $\sigma_m = 5i$ mS, and Figure 3b is calculated under $a = 0.3 \lambda$, where $d$ equals 5 mm for both cases. The dotted green lines in the two figures represent their intersection position. To investigate the achievable range of the effective permittivity, some necessary conditions must be imposed on the parameters of the HMMs:

(i) $\varepsilon_e < \varepsilon_r$; the imaginary part of the effective conductance $\sigma_m$ is negative for TM mode SISSP. So considering the relation $\varepsilon_e = \varepsilon_r - j\frac{\sigma_m(\omega, \varepsilon_r)}{\omega\varepsilon_0 d}$, the condition $\varepsilon_e < \varepsilon_r$ must be fulfilled.

(ii) $\varepsilon_\perp > 0$ $\varepsilon_{//} < 0$; here, we only consider the construction of HMMs whose effective permittivity components have opposite signs.

(iii) $a < a_{max}$. in this work, we set an upper limit on the thickness of the HMMs.

Combining all of the conditions listed above, the final achievable range of the effective permittivity are

$$\varepsilon_{/\!/} < \frac{\left(\lambda/2a_{\max}\right)^2 \varepsilon_\perp}{\varepsilon_r - \varepsilon_\perp}, \tag{22a}$$

$$\varepsilon_{/\!/} < \varepsilon_\perp. \tag{22b}$$

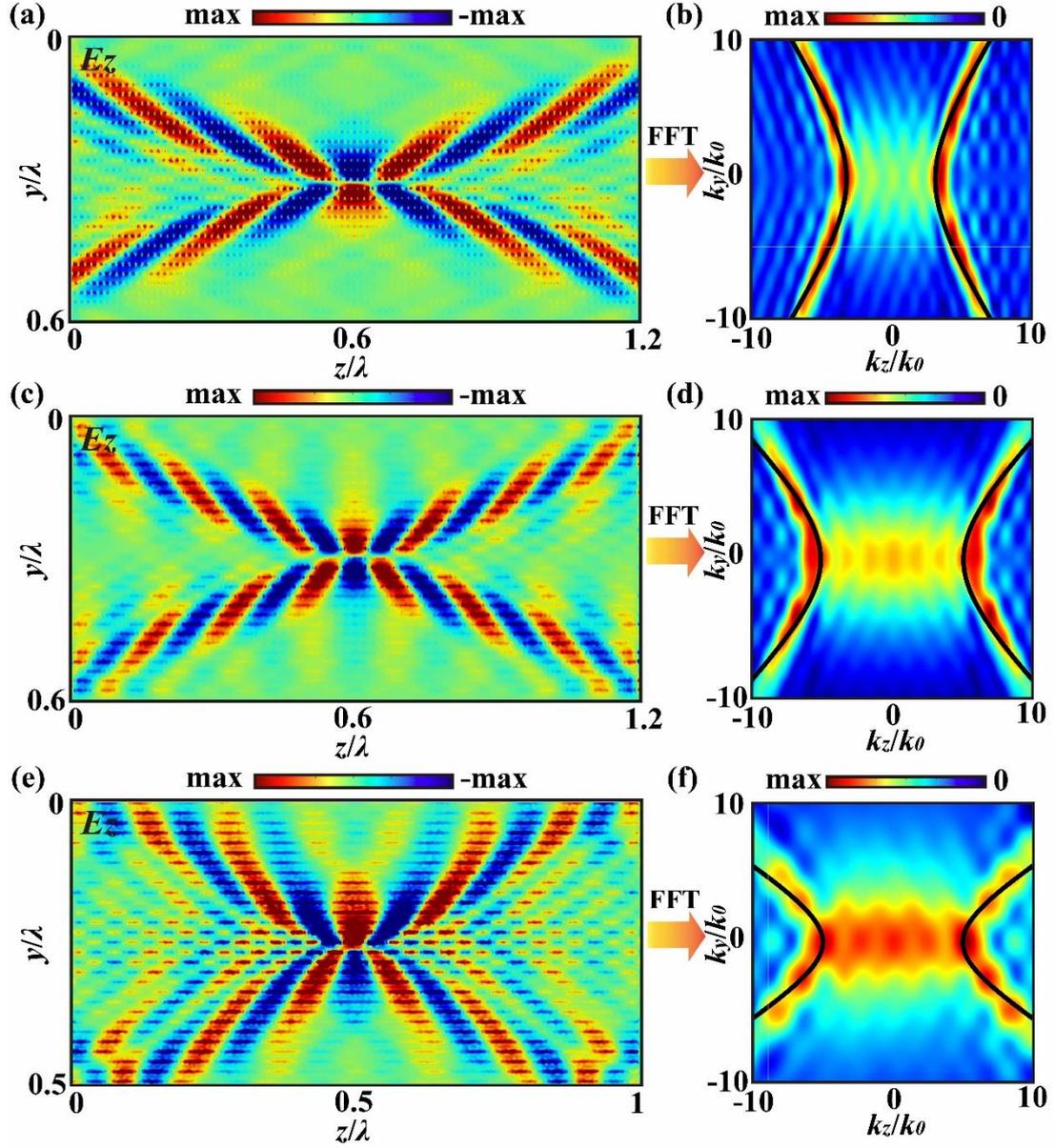

Figure 4. (a), (c), and (e) are the E$_z$ distribution of the designed HMMs with effective permittivity tensors diag (10, −25), and diag (25, −25), and diag (25, −10), respectively. The pseudo-color images in (b), (d), and (f) are the 2D spatial FFT of the corresponding E$_z$ distribution. The black lines which match well with the spatial FFT are the respective theoretical EFCs.

To intuitively visualize eqs 22a and 22b, we mark the achievable area with shading in the 2D coordinate system in Figure 3c. The corresponding asymptotic lines in the directions of two axes can also be obtained from eq 22a

$$\varepsilon_{//} = -\left(\lambda/2a_{max}\right)^2, \tag{23a}$$

$$\varepsilon_{\perp} = \varepsilon_r. \tag{23b}$$

As can be seen from Figure 3c, the range of achievable effective permittivity mainly depends on $a_{max}$ and $\varepsilon_r$. Even so, the influence is limited because the unachievable area is small enough compared with that of the achievable one.

The definition $\varepsilon_e = \varepsilon_r - j\frac{\sigma_m(\omega,\varepsilon_r)}{\omega\varepsilon_0 d}$ tells us $\sigma_m$ and $d$ have the same positions in determining $\varepsilon_e$. Comparing these two parameters, it's evident that the adjustment of distance $d$ is easier to realize. However, considering the precondition that the distance between the adjacent WPMs is deep subwavelength, the adjustment of $d$ is conditional. To investigate its influence on the dispersion characteristic, we plot the practical and theoretical EFCs of the HMM versus different $d$ under the situation $\varepsilon_r = 4$, $a = 0.3\ \lambda$, and $\sigma / d = -1i$ S / m in Figures 3d and 3e, respectively. The theoretical effective permittivity of this HMM is also determined as $\varepsilon_{\perp} = 6.2$, and $\varepsilon_{//} = -7.8$. Figure 3f displays the 2D EFCs corresponding to the cross sections in Figures 3d and 3e with $d = 2$mm, 5 mm, and 8 mm. The shadings mark the approximately consistent areas between the theoretical and practical results under different $d$. Combining all of the results above, we can conclude that the smaller the distance $d$, the better the equivalent effect of EFCs with theory. In the practical scenario, the selection of $d$ mainly depends on the required accuracy of the EFCs.

To examine the effects of the proposed method, next, we design the practical HMMs with different effective permittivity tensor diag ($\varepsilon_{\perp}$, $\varepsilon_{//}$). Three specific scenarios with diag (10, - 25), diag (25, - 25), and diag (25, - 10) which meet the conditions $\varepsilon_{\perp} \ll |\varepsilon_{//}|$, $\varepsilon_{\perp} = |\varepsilon_{//}|$, and $\varepsilon_{\perp} \gg |\varepsilon_{//}|$, respectively, are selected to demonstrate the advantage of large effective permittivity coverage of the method. Here, the central

**Table 1**

| diag($\varepsilon_\perp$, $\varepsilon_\parallel$) | $\varepsilon_r$ | $a$(mm) | $d$(mm) | $\sigma_m$ (mS) | $p_z$(mm) | $w_x$ (mm) | $w_z$(mm) | $b$(mm) |
|---|---|---|---|---|---|---|---|---|
| (10, −25) | 4.4−0.01i | 20 | 3 | -5.15i | 4 | 1.75 | 1.5 | 11.75 |
| (25, −25) | 4.4−0.01i | 16.5 | 3 | -2.9i | 3 | 1 | 1 | 12 |
| (25, −10) | 4.4−0.01i | 26.1 | 2.5 | -1.65i | 2.5 | 3.5 | 1 | 17 |

working frequency of the HMMs is $f$ = 2 GHz. For sake of demonstrating the material-independent property, the real relative permittivity of the filling materials of all the HMMs is $\varepsilon_r$ = 4.4 − 0.01i. Besides, the parameters $a$, $d$, and $\sigma_m$ can be determined by substituting the target effective permittivity into eqs 21a and 21b. The method of obtaining the exact structure parameters of the meander-line PMs with specific $\sigma_m$ can be found in Supporting Information. Table 1 lists the complete parameters of the three HMMs. To evaluate their functionalities, the Commercial FEM software COMSOL Multiphysics 5.4 is employed to calculate the field distribution. Here, a magnetic dipole with dipole moment $\hat{p} = \hat{x}$ is placed at the center of the HMMs. To reduce the reflection and imitate the infinite space, scattering boundary conditions are set around the HMMs to absorb the outgoing waves. Figures 4a, 4c, and 4e are the simulated $E_z$ distribution of the designed HMMs with effective permittivity diag ($\varepsilon_y$, $\varepsilon_z$) = diag (10, −25), and diag (25, −25), and diag (25, −10), respectively. As we can see from the figures, the visual directional propagation which is the unique phenomenon of HMMs is confirmed. Considering that all of the EMW components have approximately the same field distribution in the HMMs, here we only display the $E_z$ component. Next, to quantitatively evaluate the performance of the designed HMMs, 2D spatial fast Fourier transform(FFT) are applied on the $E_z$ components to obtain the corresponding EFCs, the results of which are illustrated by the pseudo-color image in Figures 4b, 4d, and 4e. The black lines in the spatial spectrograms are their respective theoretical EFCs ranging from −10$k_0$ to 10$k_0$. The approximate coincidence between the theoretical and practical EFCs effectively confirms the ideal effect of the proposed method. In addition to the overlapping parts with the theoretical ones, there is also some noise in the center of the spatial spectrogram. These low-frequency noises are attributed to the reflection of the imperfect boundaries, the not fine enough grid of the models in simulations, and the imperfect field distribution of the magnetic dipoles.

## Experimental Verification

To experimentally evaluate the validity of the proposed method, we fabricate the prototypes of three designed HMMs. The pictures of the practical HMMs which have the same structure parameters as the simulation ones with effective permittivity diag (10, − 25), diag(25, − 25), and diag(25, − 10) are illustrated in Figures 5a, 5c, and 5d, respectively. Their respective size is 240 mm × 30 mm × 24 mm, 180 mm × 30 mm × 20.5 mm, 150 mm × 30 mm × 30.1 mm. The samples are constructed by stacking the meander-line PMs periodically, with the oxygen-free copper plates covering the top and bottom. Figure 5b show the pictures of the meander-line PM and the copper plate of the HMM in Figure 5d. For the single-side meander-line PM, the sheet copper with a thickness of 0.017 mm is coated on the surface of the F4B substrate by the electro gold plating. The specific size of the meander-line PMs of the HMMs with effective permittivity diag (10, – 25), diag (25, – 25), and diag (25, – 10) is 240 mm × 20 mm × 3 mm, 180 mm × 16.5 mm × 3 mm, and 150 mm × 26 mm × 2.5 mm, respectively. The tiny bending parts with 90 degrees on both edges of the copper plates are used to fix the stacking meander-line PMs, as illustrated by Figure 5b.

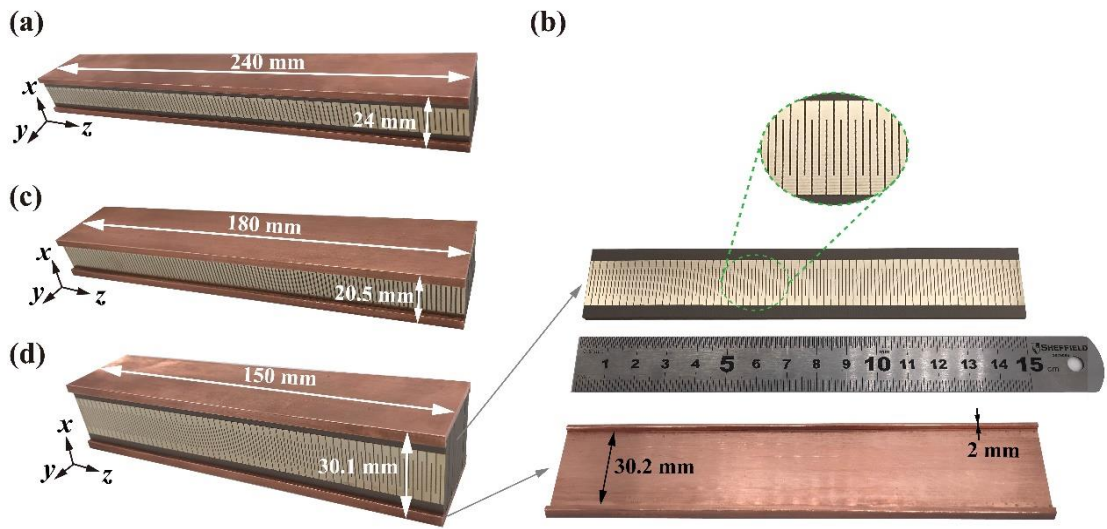

Figure 5. Pictures of the fabricated HMMs with effective permittivity (a) diag (10, − 25), (c) diag (25, − 25), and (d) diag (25, − 10). (b) The pictures of the meander-line PMs and the copper plate of (d). The inset shows the exact shape of the meander lines.

For the experimental investigation, one monopole antenna placed in front of the HMM with its probe parallel to the façade is used to excite the prototype. The other antenna is placed behind the HMMs to measure the response. The two monopole antennas connecting with Vector Network Analyzer(VNA) N5245A are fixed by the 2D electric-controlled stage. Here, the scanning step displacement of the receiving monopole antenna and the distance between the monopole antennas and the HMMs are both 2 mm. To imitate an infinite space and achieve ideal isolation between the radiating and receiving regions, the absorbing materials are placed around the HMMs.

Figures 6d, 6e, and 6f show the respective numerical $E_y$ distribution on the middle *yoz* plane of the HMMs with effective permittivity diag (10, – 25), diag (25, – 25), and diag (25, – 10), corresponding to the practical experimental configuration. The white dotted line represents the path the receiving antenna goes along to measure the response of the HMMs, and the white arrows represent the radiating monopole antennas. The white solid lines represent the boundaries between the HMMs and air. Figure 6a, 6b, and 6c display the measured and numerical normalized intensity, where the two peaks in each graph are attributed to the directional propagation in the HMMs. The approximately consistent peak distributions indicate that the EMWs generated by the radiating source propagate in the specified directions in the HMMs. The slight deviations between the simulated and experimental results are partly due to the tolerances of fabrication. Besides, the limited absorbing ability of the plane absorbing materials at 2 GHz also results in undesired reflections at both ends (*z* direction) of the HMMs. Meanwhile, the very close distance (30 mm) between the radiating and receiving regions makes it difficult to achieve perfect isolation. The EMWs leakage will cause interference to the

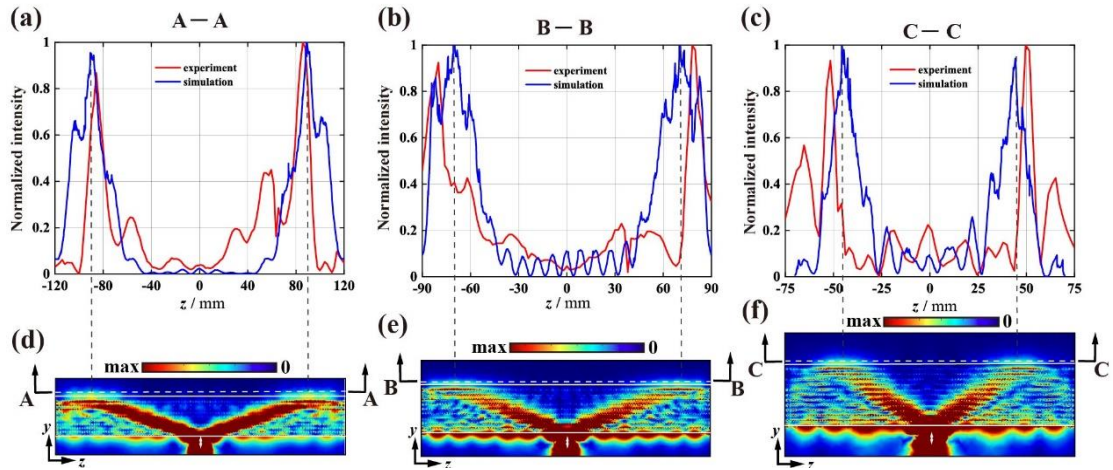

Figure 6. (a), (b), and (c) are the measured and simulated field intensity of the fabricated HMMs with effective permittivity diag (10, −25), diag (25, −25), and diag (25, −10), respectively. (d), (e), and (f) are the respective numerical field distributions of the HMMs excited by a monopole antenna. The white solid lines represent the boundaries between the HMMs and air.

measuring results, as can be seen from the side lobes at both ends of the graph in Figure 7c. The influence of these factors can be reduced by fabricating larger HMMs and using better absorbing materials. Combing all of the results, we can conclude that the validity of the designed HMMs is fairly verified.

## Design of Air-medium HMM

To further illustrate the superiority of the proposed method, we design a novel air-medium HMM which is mainly made of copper. The all-metal structure can obviously reduce the dielectric loss compared with the conventional HMMs because air can be considered as lossless. This advantage is more evident and valuable in the terahertz where the loss of dielectric is quite higher. Besides, the absence of dielectric makes the air-medium HMM qualified to work in some specific scenarios such as low-threshold or non-threshold Cherenkov radiation in which the electrons should move freely. [39-41] The schematic of the designed air-medium HMM whose central working frequency is $f$ = 1GHz is illustrated in Figure 7a, where the inset shows its front view. The core part of the HMM is only composed of meander-line PMs and copper plates. The additional parts aiming at fixing the unsupported meander-line PMs are added on both sides of the core part. The effective permittivity of the air-medium HMM is diag (6, − 12), and the corresponding structure parameters of the core part are $w_x$ = 1 mm, $w_z$ = 1 mm, $b$ = 28 mm, $p_z$ = 4 mm, $a$ = 47.37 mm, and $d$ = 7 mm. Besides, the length ($z$ direction) and width ($y$ direction) of the core part is 252 mm and 154 mm, respectively. In order to effectively fix the meander lines without affecting the normal functions, the additional parts with approximately the same impedance as the core part are designed. The resultant parameters of the additional parts with effective permittivity diag (7.2, − 14.4) are $w_x$ = 1 mm, $w_z$ = 3 mm, $b$ = 28 mm, $p_z$ = 8 mm, $a$ = 47.37 mm, $\varepsilon_r$ = 2.2 − 0.01$i$ and $d$ = 7 mm. Besides, the length ($z$ direction) of each additional part is 24 mm, which is small enough not to affect the normal functions of the core part. To examine the practical performance of the air-medium HMM, we numerically calculate its field distribution under the excitation a magnetic dipole with dipole moment $\hat{p} = \hat{x}$. Figure 7c illustrates the $E_z$ distribution on the middle plane of the HMM. The dotted lines in the figure represent the boundary between the core and additional parts. As can be seen from the figure, there is no evident reflection between the boundaries of the two parts, indicating the perfect effect of the impedance match. Besides, the practical EFC calculated by the spatial FFT of the $E_z$ distribution in Figure 7c also matches well with the theoretical one, which effectively demonstrates the functions of the air-medium HMMs.

To quantitatively evaluate the loss of the designed HMM, we numerically calculate its damping in time versus spatial harmonics under different loss tangent. In fact, the air is a lossless dielectric. However, it is not fair to directly compare the air-medium HMMs with those filled with the lossy materials because of the different relative permittivity. Therefore, for the convenience of evaluating the performance of the air-medium HMM, we add the virtual loss on air. In the practical simulation, we calculate the Eigen frequency and damping in time by adding the periodic boundary conditions with specified $k_z$ and $k_y$ in the $z$ and $y$ directions of the unit cell of the HMM. As illustrated by Figure 7b, the calculated Eigen frequency, which should theoretically be strictly equal to 1GHz for all the specified wave vector configuration, is actually distributed in a narrow frequency range around 1GHz. The results can be optimized by reducing the interval $d$ between the adjacent WPMs. Meanwhile, we also plot the damping in time of the air-medium HMMs under $\tan\delta = 0$, 0.002, and 0.004 in Figure 7b, from which we find the propagation loss increases with the increase of $\tan\delta$. Besides, comparing the real air with the situation where $\tan\delta = 0.004$, the total loss of the HMM reduces by nearly half, demonstrating the superiority of the air-medium HMM in low-loss scenarios.

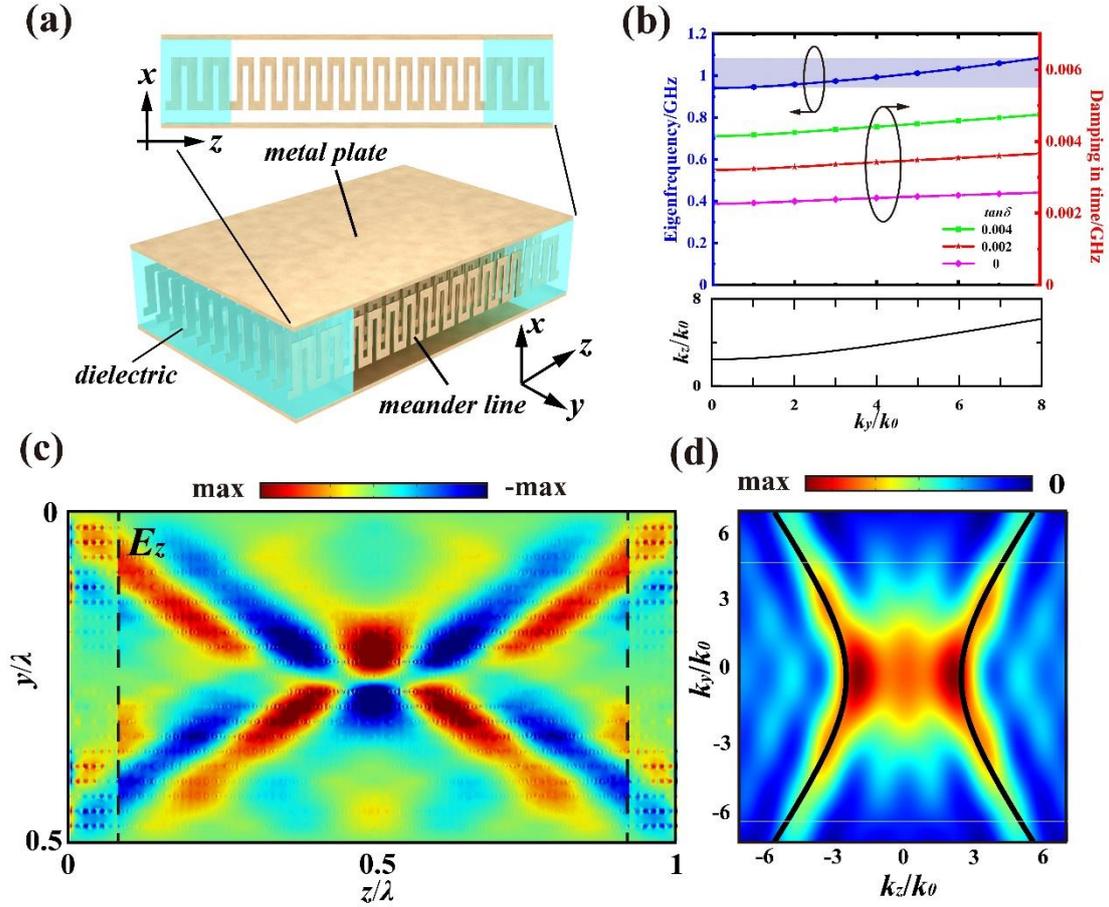

Figure 7. (a) Schematic of the air-medium HMM, where the inset shows its front view. (b) The Eigen frequency and the damping time of the HMM versus spatial harmonics under different loss tangent. (c) The electric field $E_z$ distribution on the middle plane of the air-medium HMMs excited by a magnetic dipole. (d) The spatial FFT of the $E_z$ distribution in (c). The black lines represent the theoretical EFCs of the designed HMM.

## Discussion of Potential Applications

Due to the superiority of materials independence and large permittivity coverage, the proposed method has a wide range of potential applications from conventional focus and super-resolution imaging to electromagnetic cloaking. As proofs, we discuss the feasibility and advantages of designing ENZ materials and super-resolution lenses based on the proposed method.

The gray area in Figure 3c displays the achievable effective permittivity range of the HMMs constructed by the WPMs. However, when we expand the scope to the whole permittivity space, the coordinate origin which corresponds to ENZ materials also belongs to the achievable area. When we elaborately design the meander-line PMs so that its effective conductivity $\sigma_m/d = -j\omega\varepsilon_0(\varepsilon_r - (\lambda/2a)^2)$, the orthogonal components of the effective permittivity tensors both equal zero. As can be seen from the previous analysis, this specified effective conductivity is not difficult to achieve. In this way, ENZ materials can be successfully obtained. Compared with other approaches, the proposed method also has no strict demand on the compositional materials. It should be noted that the effective conductance of the meander-line PMs with a larger imaginary part is always more difficult to design. Therefore, although arbitrary medium can be selected, the filling materials with a relatively large permittivity can help reduce the difficulty in design.

The hyperlenses which can realize subwavelength imaging by allowing the propagation of both traveling and evanescent waves can also be designed based on the proposed method. Different from that of the conventional HMMs, the negative permittivity components of the hyperlenses approach negative infinity. This characteristic puts an extremely high demand on its design. Here, according to eqs 21 and S10 (Section 2, Supporting Information), we know that the negative infinity permittivity $\varepsilon_{//}$ can be obtained by designing the meander-line PMs with the propagation constant $k_z$ slightly larger than $\sqrt{\varepsilon_r}k_0$. Under this condition, the positive components of the effective permittivity are approximately equal $\varepsilon_r$ and are independent of $a$. In this way, a novel material-independent hyperlens is also successfully obtained. Here, the ultimate spatial resolution is mainly determined by the distance between the adjacent WPMs.

## Conclusion

In this work, we propose the concept of SISSP supported by WPMs and prove that hyperbolic dispersion can be obtained by stacking the WPMs periodically. The proposed method completely breaks the barriers of materials, and arbitrary materials can be selected to construct the HMMs with independent effective permittivity. Three proof-of-concept HMMs are designed and with their functions validated numerically and experimentally by directional propagation. Besides, a novel air-medium HMM that has extremely small damping in time is also designed and demonstrated. It should be noted that this method can also be extended to terahertz wavelengths. This work may hold promise in designing microwave and terahertz epsilon-near-zero materials, superlens, focusing lenses, and so on.

## Acknowledgements

This work was supported by National Key Research and Development Program under Grant No. 2019YFA0210203 and National Natural Science Foundation of China under Grant No. 61971013.

**Conflict of Interest**: The authors declare no conflict of interest.

## References:

1. Jacob, Z.; Alekseyev, L. V.; Narimanov, E. Optical hyperlens: far-field imaging beyond the diffraction limit. *Opt. Express* **2006**, *14* (18), 8247-8256.
2. Lee, H.; Liu, Z.; Xiong, Y.; Sun, C.; Zhang, X. Development of optical hyperlens for imaging below the diffraction limit. *Opt. Express* **2007**, *15* (24), 15886-15891.
3. Liu, Z.; Lee, H.; Xiong, Y.; Sun, C.; Zhang, X. Far-field optical hyperlens magnifying sub-diffraction-limited objects. *Science* **2007**, *315* (5819), 1686-1686.
4. Lu, D.; Liu, Z. Hyperlenses and metalenses for far-field super-resolution imaging.

*Nat. Commun.* **2012**, *3* (1), 1-9.

5. Andryieuski, A.; Lavrinenko, A. V.; Chigrin, D. N. Graphene hyperlens for terahertz radiation. Physical Review B **2012**, *86* (12), 121108.

6. Argyropoulos, C.; Estakhri, N. M.; Monticone, F.; Alù, A. Negative refraction, gain and nonlinear effects in hyperbolic metamaterials. *Opt. Express* **2013**, *21* (12), 15037-15047.

7. Bang, S.; So, S.; Rho, J. Realization of broadband negative refraction in visible range using vertically stacked hyperbolic metamaterials. *Sci. Rep.* **2019**, *9* (1), 1-7.

8. García-Chocano, V. M.; Christensen, J.; Sánchez-Dehesa, J. Negative refraction and energy funneling by hyperbolic materials: An experimental demonstration in acoustics. *Phys. Rev. Let.* **2014**, *112* (14), 144301.

9. Sreekanth, K.; De Luca, A.; Strangi, G. Negative refraction in graphene-based hyperbolic metamaterials. *Appl. Phys. Lett.* **2013**, *103* (2), 023107.

10. Desouky, M.; Mahmoud, A. M.; Swillam, M. A. Tunable mid IR focusing in InAs based semiconductor hyperbolic metamaterial. *Sci. Rep.* **2017**, *7* (1), 1-7.

11. Kannegulla, A.; Cheng, L.-J. Subwavelength focusing of terahertz waves in silicon hyperbolic metamaterials. *Opt. Lett.* **2016**, *41* (15), 3539-3542.

12. Noginov, M.; Lapine, M.; Podolskiy, V.; Kivshar, Y. Focus issue: hyperbolic metamaterials. *Opt. Express* **2013**, *21* (12), 14895-14897.

13. Sedighy, S. H.; Guclu, C.; Campione, S.; Amirhosseini, M. K.; Capolino, F. Wideband planar transmission line hyperbolic metamaterial for subwavelength focusing and resolution. *IEEE Trans. Microwave Theory Tech.* **2013**, *61* (12), 4110-4117.

14. Guo, Y.; Cortes, C. L.; Molesky, S.; Jacob, Z. Broadband super-Planckian thermal emission from hyperbolic metamaterials. *Appl. Phys. Lett.* **2012**, *101* (13), 131106.

15. Guo, Y.; Jacob, Z. Thermal hyperbolic metamaterials. *Opt. Express* **2013**, *21* (12), 15014-15019.

16. Galfsky, T.; Krishnamoorthy, H.; Newman, W.; Narimanov, E.; Jacob, Z.; Menon, V. Active hyperbolic metamaterials: enhanced spontaneous emission and light extraction. *Optica* **2015**, *2* (1), 62-65.

17. Lu, D.; Kan, J. J.; Fullerton, E. E.; Liu, Z. Enhancing spontaneous emission rates of molecules using nanopatterned multilayer hyperbolic metamaterials. *Nat. nanotechnol.* **2014**, *9* (1), 48-53.

18. Sreekanth, K.; Biaglow, T.; Strangi, G. Directional spontaneous emission enhancement in hyperbolic metamaterials. *J. Appl. Phys.* **2013**, *114* (13), 134306.

19. Tumkur, T.; Zhu, G.; Black, P.; Barnakov, Y. A.; Bonner, C.; Noginov, M. Control of spontaneous emission in a volume of functionalized hyperbolic metamaterial. *Appl. Phys. Lett.* **2011**, *99* (15), 151115.

20. Bergman, D. J. The dielectric constant of a composite material—a problem in classical physics. *Phys. Rep.* **1978**, *43* (9), 377-407.

21. Rytov, S. Electromagnetic properties of a finely stratified medium. *Sov. Phys. JEPT* **1956**, *2*, 466-475.

22. Wood, B.; Pendry, J.; Tsai, D. Directed subwavelength imaging using a layered metal-dielectric system. *Phys. Rev. B* **2006**, *74* (11), 115116.

23. Chang, Y.-C.; Liu, C.-H.; Liu, C.-H.; Zhang, S.; Marder, S. R.; Narimanov, E. E.; Zhong, Z.; Norris, T. B. Realization of mid-infrared graphene hyperbolic metamaterials. *Nat. Commun.* **2016**, *7* (1), 1-7.

24. Dai, S.; Ma, Q.; Liu, M.; Andersen, T.; Fei, Z.; Goldflam, M.; Wagner, M.; Watanabe, K.; Taniguchi, T.; Thiemens, M. Graphene on hexagonal boron nitride as a tunable hyperbolic metamaterial. *Nat. nanotechnol.* **2015**, *10* (8), 682-686.

25. Iorsh, I. V.; Mukhin, I. S.; Shadrivov, I. V.; Belov, P. A.; Kivshar, Y. S. Hyperbolic metamaterials based on multilayer graphene structures. *Phys. Rev. B* **2013**, *87* (7), 075416.

26. Othman, M. A.; Guclu, C.; Capolino, F. Graphene-based tunable hyperbolic metamaterials and enhanced near-field absorption. *Opt. Express* **2013**, *21* (6), 7614-7632.

27. Zhang, X.; Liu, Z. Superlenses to overcome the diffraction limit. *Nat. Mater.* **2008**, *7* (6), 435-441.

28. Grigorenko, A.; Polini, M.; Novoselov, K. Graphene plasmonics. *Nat. Photonics* **2012**, *6* (11), 749-758.

29. Della Giovampaola, C.; Engheta, N. Plasmonics without negative dielectrics. *Phys. Rev. B* **2016**, *93* (19), 195152.

30. Li, Z.; Liu, L.; Sun, H.; Sun, Y.; Gu, C.; Chen, X.; Liu, Y.; Luo, Y. Effective surface plasmon polaritons induced by modal dispersion in a waveguide. *Phys. Rev. Appl.* **2017**, *7* (4), 044028.

31. Wang, C.; Wang, H.; Shen, L.; Abdi-Ghaleh, R.; Musa, M. Y.; Xu, Z.; Zheng, B. Structure-Induced Hyperbolic Dispersion in Waveguides. *IEEE Trans. Antennas Propag.* **2019**, *67* (8), 5463-5468.

32. Huang, T.-J.; Yin, L.-Z.; Zhao, J.; Du, C.-H.; Liu, P.-K. Amplifying Evanescent Waves by Dispersion-Induced Plasmons: Defying the Materials Limitation of the Superlens. *ACS Photonics* **2020**, *7* (8), 2173-2181.

33. Ji, W.; Zhou, X.; Chu, H.; Luo, J.; Lai, Y. Theory and experimental observation of hyperbolic media based on structural dispersions. *Phys. Rev. Mater.* **2020**, *4* (10), 105202.

34. Yin, L.-Z.; Huang, T.-J.; Wang, D.; Liu, P.-K. Hyperbolic spoof plasmons in layered equivalent graphene metasurfaces. arXiv preprint arXiv:2006.12255 2020.

35. Li, Z.; Sun, Y.; Wang, K.; Song, J.; Shi, J.; Gu, C.; Liu, L.; Luo, Y. Tuning the dispersion of effective surface plasmon polaritons with multilayer systems. *Opt. Express* **2018**, *26* (4), 4686-4697.

36. Prudêncio, F. R.; Costa, J. R.; Fernandes, C. A.; Engheta, N.; Silveirinha, M. G. Experimental verification of 'waveguide' plasmonics. *New J. Phys.* **2017**, *19* (12), 123017.

37. Li, Y.; Zhang, Z. Experimental verification of guided-wave lumped circuits using waveguide metamaterials. *Phys. Rev. Appl.* **2018**, *9* (4), 044024.

38. Gomez-Diaz, J. S.; Tymchenko, M.; Alu, A. Hyperbolic plasmons and topological transitions over uniaxial metasurfaces. *Phys. Rev. Lett.* **2015**, *114* (23), 233901.

39. Liu, F.; Xiao, L.; Ye, Y.; Wang, M.; Cui, K.; Feng, X.; Zhang, W.; Huang, Y. Integrated Cherenkov radiation emitter eliminating the electron velocity threshold. *Nat. Photonics* **2017**, *11* (5), 289.

40. Su, Z.; Xiong, B.; Xu, Y.; Cai, Z.; Yin, J.; Peng, R.; Liu, Y. Manipulating Cherenkov

radiation and smith–purcell radiation by artificial structures. *Adv. Opt. Mater.* **2019**, *7* (14), 1801666.

41. Tao, J.; Wu, L.; Zheng, G.; Yu, S. Cherenkov polaritonic radiation in a natural hyperbolic material. *Carbon* **2019**, *150*, 136-141.